\documentclass{jaa}
\usepackage{natbib}
\usepackage{longtable}
\usepackage{pdflscape}
\usepackage{multirow}
\usepackage{url}
\usepackage{hyperref}
\usepackage{longtable}
\usepackage{lineno}
\usepackage{graphicx}
\usepackage{lscape}
\usepackage{pdflscape}
\begin{document}\sloppy

%%paper title
%%For line breaks \\ can be used within title
\title{Propagation Characteristics of the April 21, 2023 CME}
%\linenumbers

%%author names are separated by comma (,)
%%use \and before the last author name
%%use a * along with the number separated by comma
%% for the  author for correspondence
%%\textsuperscript{number} is used for affiliation
%%\affilOne, \affilTwo etc., upto \affilTwentyfive is possible
%%Please note the first letter after \affil is capitalised in the command
%%
%
%\author{Sandeep Kumar \textsuperscript{1,2}}

%\author{Nandita Srivastava\textsuperscript{2}}

\author{ Sandeep Kumar\textsuperscript{1,2,*}, Nandita Srivastava\textsuperscript{2}, Parthib Banerjee \textsuperscript{3}, Nat Gopalswamy \textsuperscript{4}}
\affilOne{\textsuperscript{1}Discipline of Physics, Indian Institute of Technology Gandhinagar, Palaj, Gandhinagar-382 355, Gujarat, India\\}
\affilTwo{\textsuperscript{2}Udaipur Solar Observatory, Physical Research Laboratory, Udaipur, 313001, India\\}
\affilThree{\textsuperscript{3}Indian Institute of Science Education and Research (IISER) Berhampur,760003, Odisha ,India. \\}
\affilFour{\textsuperscript{4}NASA Goddard Space Flight Center, Greenbelt, MD 20771, USA \\}

%%escape two column mode for title, affiliation and abstract
%%by giving \twocolumn command as shown

\twocolumn[{

\maketitle

%%include \corres to print the corresponding author Email id
\corres{ssandeepsharma201@gmail.com}

%%include \msinfo for
%%manuscript information such as
%%received, revised and accepted dates
%%
%\msinfo{31 January 2025}{31 January 2025}

%%abstract
\begin{abstract}
Accurate estimation of propagation characteristics of coronal mass ejections (CMEs) is crucial for predicting their geoeffectiveness. Stereoscopic techniques to study the kinematics of CMEs generally have been carried out using remote sensing observations from three viewpoints, i.e. STEREO-A, STEREO-B, and SOHO. Since the loss of STEREO-B in 2014, stereoscopic reconstruction of CMEs has been restricted to the observations from only two viewpoints, i.e., STEREO-A and SOHO. When the angle of separation between STEREO-A and SOHO is small, it leads to larger uncertainties in the CME kinematics derived using stereoscopic techniques. In this paper, we demonstrate how this limitation can be addressed and how uncertainties in the estimation of CME kinematics and propagation direction can be reduced. For this purpose, we selected the CME of April 21, 2023, which was observed by two spacecraft, i.e. STEREO-A and SOHO, separated by a small $10^\circ$ angle. Using the Graduated Cylindrical Shell (GCS) model on the remote-sensing observations near the Sun and the Advanced Drag-Based Model (ADBM) in the heliosphere, we estimated the arrival time of the CME at different locations in the heliosphere and compared it with the actual arrival time obtained from the in-situ measurements taken by three spacecraft, BepiColombo, STEREO-A and Wind. Our analysis reveals a directional uncertainty of approx $20 ^\circ$ from observations from two viewpoints. These uncertainties significantly affect the
arrival-time prediction of the CME. We consider the actual chronology of CME arrival times at STEREO-A and Wind as critical parameters to constrain the direction of propagation, which serves as a key input in the ADBM. The chronology of arrival of the CME ejecta at STEREO-A, which is 4.5  hrs earlier than at Wind, proved essential for resolving directional ambiguities in the GCS reconstruction model.
\end{abstract}

%%insert keywords separated by 3 hyphens using \keywords{words}
}]
%%close the twocolumn escape here

%%include \doinum{number}for the DOI number in the header
%%include \volnum{number} for the volume number in the header
%%include \year{yyyy} for  year of publication in the header
%%include \pgrange{num--num} page range of article in the header
%%include \artcitid{num} for the article citation id
%%include \lp to print last page of the article
%%include \setcounter{page}{pagenum} for the exact starting page of the article

\doinum{12.3456/s78910-011-012-3}
\artcitid{\#\#\#\#}
\volnum{000}
\year{0000}
\pgrange{1--}
\setcounter{page}{1}
\lp{1}

\section{Introduction}
Coronal mass ejections (CMEs) are energetic eruptive events from the solar corona. Earth-directed CMEs interact with magnetosphere of the Earth and CME with a southward (negative Bz) interplanetary magnetic field (IMF) can cause a geomagnetic storm that results in adverse space weather effects. CMEs can deflect and rotate in the heliosphere due to various interactions, including those with other CMEs, stream interacting regions, the ambient solar wind, and the coronal and interplanetary magnetic field \citep{Wang:2004,Gopalswamy:2009,gui:2011,Wang:2014,Kay:2015,Kay:2017,kumar:2023,kumar:2024}. These interactions complicate the estimation of CME properties at L1 based on their near-Sun observations. Therefore, studying CME propagation in the heliosphere is crucial for accurately determining and predicting their geoeffectiveness on Earth.
The geoeffectiveness of a CME is determined by several factors, including the strength of the southward component (negative Bz) of the IMF, density, velocity, and the duration of negative Bz.  \citep{tang:1989,gos:1990,echer:2008}. The negative Bz component interacts with the magnetopause, causing magnetic reconnection, which causes the exchange of energy between the CME and the magnetosphere of Earth \citep{dungey:1961,gonzalez:1974,gonzalez:1999} and thus creates a geomagnetic storm that causes a decrease in the Dst Index. Intense geomagnetic storms having Dst ($\le$-100 nT) can cause serious space weather consequences \citep{tsurutani:1997,zhang:2003,srivastava:2004,ex:2014,gopalswamy:2024}. Thus, studying these events and improving space weather prediction becomes essential.
Apart from studying the properties of the geo-effectiveness of CMEs, there is a need to make an arrival time prediction model for them and obtain their transit speed to minimize the potential damage caused by the CMEs. Solar and Heliospheric Observatory (SOHO) \cite{domingo:1995}, the Large Angle Spectrometric Coronagraph (LASCO) \cite{brueckner:1995} routinely observes the CMEs. ACE \citep{stone:1998} and Wind spacecraft \citep{ogilive:1995} measure the in-situ properties of the CMEs at 1 A.U. \cite{mishra:2021} recently reported that in-situ properties of an ICME may differ at different longitudinal locations in the heliosphere. Therefore, it is crucial to have a multiple in-situ observation of ICMEs at different heliocentric points. Thus, in-situ and remote-sensing observations of CMEs help to analyze the Earth-directed CME till 1 AU \citep{Gopalswamy:2001(a)}. However, due to a lack of observations between the in-situ and remote-sensing observations, it was not possible to estimate the true speed and direction of the Earth-directed CMEs. The STEREO mission \citep{kaiser:2008} has been providing multiple viewpoints observations of CMEs, enabling estimation of the true CME speeds in the coronagraphic (COR2)  and Heliospheric Imager (HI) fields-of-view  (FOV). Several 3D reconstruction models have been developed to track the 3-dimensional view of  CMEs (e.g., tie-pointing: \cite{inhester:2006}; forward modelling: \cite{thernisien:2009}; polarization ratio: \cite{moran:2004}).

Different models can be used in the heliosphere to estimate the time of arrival (ToA) of the CMEs and their impact parameters. This approach included complete 3D MHD modeling of the ambient solar wind environment with a model of the CME \citep{kumar:2020}, e.g., cone model \citep{Xie:2004}, \citep[FRi3D;][]{isavnin:2016}, and spheromak \citep{tal:2020}. Although these models can predict the in-situ properties at L1, they are computationally expensive. Given the fact that the physics of the heliosphere is mostly drag-dominated, simple models are also used in the community, which are faster as compared to the full 3D MHD models and provide a reasonably reliable estimate of the ToA of the CME at L1.

These models include Empirical CME Arrival (ECA) by \cite{gopalswamy:2001(b)}, which had an error range of 35 hrs, Empirical Shock Arrival (ESA) by  \cite{gopalswamy:2005}, which is a modified version of ECA  where a CME is considered as a driver of magnetohydrodynamics (MHD) shock and reduces the error in the arrival time to 30 hr. Various papers documented the forecasting of CME arrival time using the empirical relationship between the projected speed of the CME  and the arrival time estimation of various events, e.g., \citep{gopalswamy:2001(b),vrsnak:2002,schwenn:2005}. The analytical Drag Based Model \citep{vrsnak:2007,lara:2009,vrsnak:2010} and numerical MHD simulation model \citep{odstrcil:2004,manchester:2004,smith:2009} reduced the errors in the arrival time to 10 hrs or slightly smaller \citep{Gopalswamy:2013}. The drag-based model estimates the arrival time of the apex of the CME at Earth. Although it considers acceleration or deceleration of the CME  due to the drag, it does not consider the geometry of the CME, which was incorporated later in the advanced drag-based model \citep[ADBM;][]{vrsnak:2013}. The ADBM considers both the ambient solar wind drag and the geometry of CMEs. Recently \citep{Kay:2024} analyzed CME arrival predictions using various CME ToA prediction models at the Community Coordinated Modeling Center (CCMC) and found that well-established models typically achieve mean absolute errors of $\approx$ 10–14 hours. Errors remain strongly dependent on CME transit speed and reconstruction uncertainties.

One of the major CMEs of the solar cycle 25 was observed on 21 April 2023, which reached the Earth on 23 April 2023. This CME was a high-speed CME with a projected speed of more than 1000 km/s, leading to a strong geomagnetic storm on April 24, 2023, with a Dst index of -213 nT \citep {gagh:2024,gopalswamy:2024}. The CME originated around 18:00 UT on April 21, 2023, from the AR 13283 near the disc center S21W11,  The CME was associated with the M1.7 flare observed by GOES \citep{vemareddy:2024}. \cite{gopalswamy:2024}, used the cone-based Model \citep{gopalswamy:2015} to estimate the shock arrival of the CME and suggested a westward deflection of the CME. \cite{gagh:2024} investigated the geomagnetic storm caused by this  CME on Earth, where they found a two-step decrease in the Dst index, once during the passage of the sheath region and another during the passage of the Magnetic Cloud (MC) region of the CME. This CME produced the largest storm of SC 25 till that date. Thus, it provided a unique opportunity to understand its propagation in the heliosphere and its impact on the magnetosphere of the Earth. 

Recently \cite{verbek:2023} reported that limited viewpoints and smaller spacecraft separation angles introduce strong projection effects and parameter degeneracies in the 3D stereoscopic reconstruction process, leading to substantial uncertainties, particularly in the estimation of CME longitude, tilt, and angular width. Their results demonstrate that the inclusion of at least two well-separated viewpoints is essential to mitigate these errors, thereby improving the robustness of the derived CME parameters and enhancing the reliability of subsequent CME propagation and arrival-time forecasts. April 21, 2023, CME was observed by STEREO-A and SOHO separated by a small angle of only $10^\circ$. Therefore, this CME provides a valuable case for assessing the limitations and inherent uncertainties arising from such restricted viewing geometry.

\section{Data}

In order to track the CME of April 21, 2023, near the Sun, we used observations from the SOHO and STEREO-A spacecraft for 3D reconstruction of the  CME. Sun-Earth Connection Coronal and Heliospheric Investigation (SECCHI) on board STEREO spacecraft is a suite of five telescopes (Extreme Ultraviolet Imager (EUVI): 1–1.7$R_\odot$; COR1: 1.5–4.0$R_\odot$; COR2: 2.5–15.0$R_\odot$; HI1: 15–90 $R_\odot$; and HI2: 70–330 $R_\odot$) \citep{howard:2008} that image CMEs continuously from the Sun to the Earth and beyond. Both coronagraphs (COR1 and COR2) are pointed at the Sun while both HIs (HI1 and HI2) are off-pointed from the Sun at a solar elongation of 14° and $53.7^{\circ}$, respectively \citep{eyles:2009}. HI1 and HI2 have a wide FOV of 20° and 70°, respectively, and have their optical axes aligned in the ecliptic plane. LASCO-C2 \& C3 coronagraphs  \citep{brueckner:1995} combined with STEREO-A $\&$ B viewpoints, enabled the 3D reconstruction of the CMEs from three viewpoints. 

\begin{figure}%%[H]
\begin{center}
\includegraphics[width=\linewidth]{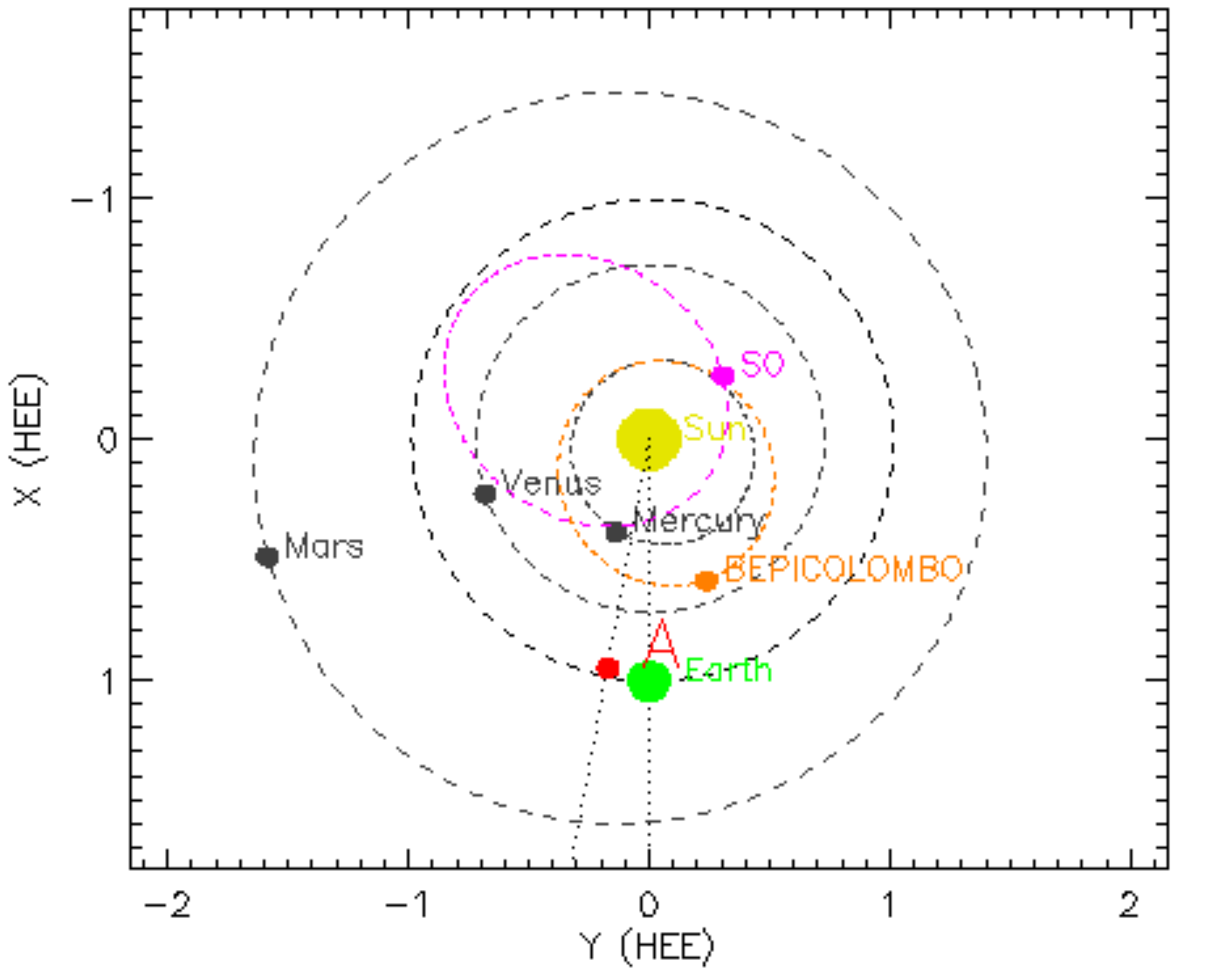}
\caption{The relative orientation and location of different spacecraft, i.e., STEREO-A (red),  BepiColombo (orange), and Earth/Wind (green) on 23 April 2023. The angular separation between Wind and STEREO-A is nearly -10°, and between Wind and BepiColombo is nearly +22°. The image was generated by the online tool of Stereo Science Center \protect \footnote{\url{https://stereo-ssc.nascom.nasa.gov/cgi-bin/make_where_gif}}.}
\label{fig:Loc}
\end{center}
\end{figure}

As the April 21, 2023, CME propagated in the heliosphere, it passed through different spacecraft located at various heliocentric distances. For this particular CME, we used the in-situ data from Wind and STEREO-A located at $\sim 0.997$ AU and $\sim 0.963$ AU, respectively. We have also referred to the CME arrival time at BepiColombo \citep{bepi:2021} located a  $\sim 0.631$ AU from the HELCATS(WP4) Catalogue \citep{hell:2020} \footnote{\url{https://www.helcats-fp7.eu/catalogues/wp4_icmecat.html}}.

Figure \ref{fig:Loc} shows the relative orientations and location of STEREO-A (red),  BepiColombo (Orange), and Earth/Wind (green) on 23 April 2023.  The information on the arrival time of different parts of the CME in the three spacecraft was obtained from the HELCATS(WP4) Catalogue \citep{hell:2020} \footnote{\url{https://www.helcats-fp7.eu/catalogues/wp4_icmecat.html}} .
%and Richardson-Cane Catalogue \footnote{(\url{https://izw1.caltech.edu/ACE/ASC/DATA/level3/icmetable2.htm})}

\section{Analysis}
\subsection{GCS reconstruction of the CME close to the Sun }
 
The projected speed of the CME in the LASCO FOV was 1284 km/s at 18:12 UT on April 21, 2023, as listed in the CDAW catalogue\footnote{\url{https://cdaw.gsfc.nasa.gov/CME_list/UNIVERSAL_ver2/2023_04/univ2023_04.html}} \citep{seiji_2004}. The associated halo CME was observed in LASCO-C2 and C3 coronagraphs from 18:54 UT to 23:54 UT and by the COR2 coronagraph between 18:54 UT and 20:08 UT. 
To derive its deprojected kinematics near the Sun, we applied a 3D reconstruction using the Graduated Cylindrical Shell (GCS) technique \citep{thernisien:2006}.  Figure \ref{fig:Loc} shows a small angular separation of $10^\circ$ between STEREO-A and Earth, which may introduce significant errors in the GCS fitting as discussed by \cite{verbek:2023}.

We tracked the CME flux rope in the HI1 fov from 20:08 UT to 23:54 UT using an extended version of the Python module, name \textit{gcs\_python}\footnote{\url{https://github.com/johan12345/gcs_python/tree/master/gcs}}, which implements GCS reconstruction in LASCO-C2 and C3, SECCHI-COR2 and HI1 images \citep{kumar:2023,gcs:2024,kumar:2024}.
The flux rope geometry in GCS model is described by six parameters: half-angle, kappa, height, latitude, longitude, and tilt. The first three parameters determine the flux rope geometry, while longitude and latitude specify its azimuthal and latitudinal positions, respectively. The tilt defines its orientation relative to the ecliptic plane.

 \begin{figure*}%%[H]
\begin{center}
\includegraphics[width=12cm]{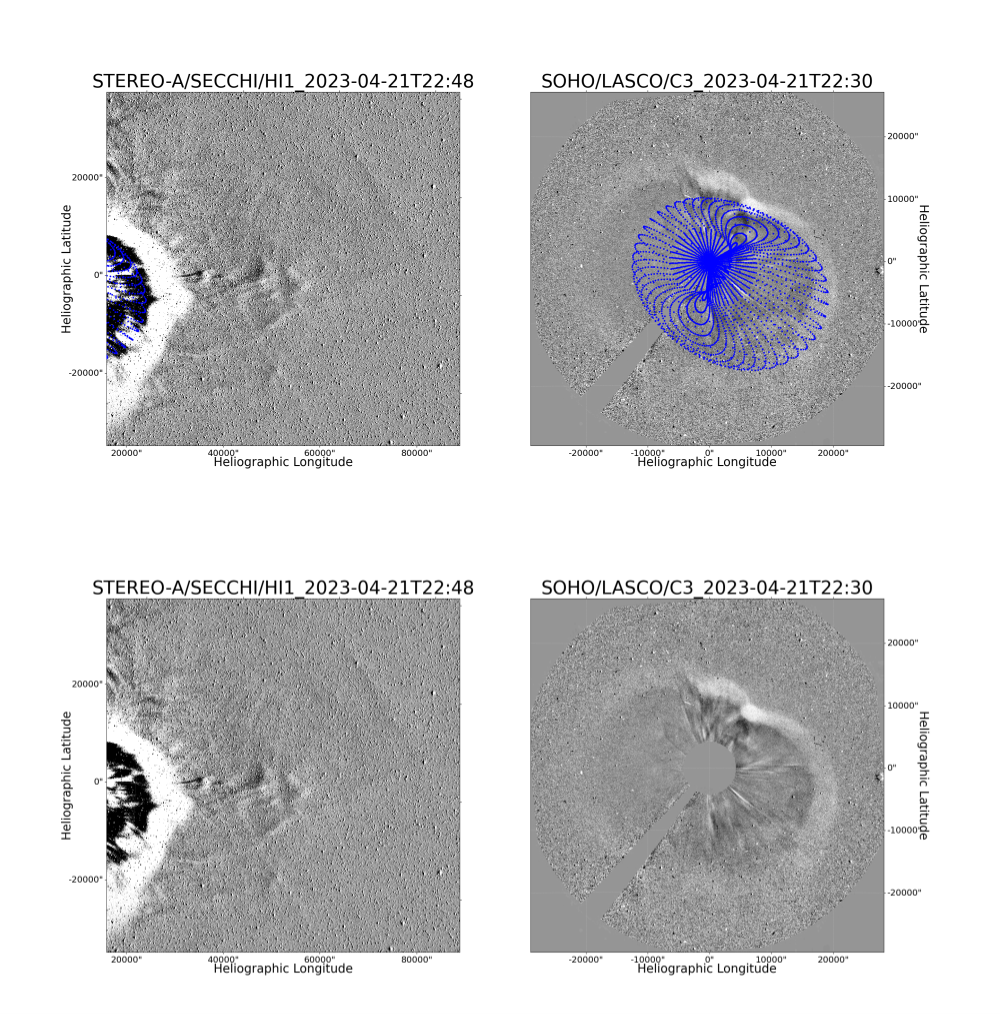}
\caption{GCS reconstruction of the 21 April 2023 CME using observations from two viewpoints of STEREO-A/HI1 and SOHO LASCO-C3. The top panel shows the CME overplotted with GCS reconstructed mesh. The bottom panel shows a CME without GCS mesh at 21 April 2023 22:30 UT.}
\label{fig:gcs}
\end{center}
\end{figure*}
Two different cases for GCS fittings were done by two of the authors, as mentioned below. 
\begin{enumerate}
    \item In the first GCS fitting of the CME on images from STEREO and SOHO viewpoints, recorded between during 18:24 to 23:54 UT, the longitude of the CME was found to remain constant (nearly +20$^\circ$) as the CME propagated from COR2 FOV to HI.
   This suggests a final $20^\circ$ westward propagation of the CME in the heliosphere at $\approx$ 38 $R_\odot$.
   
    \item In the second fitting, we observed a decrease in longitude(+34 to +0$^\circ$.) as the CME propagated from COR2 FOV to HI, it corresponds to deflection towards the East, suggesting a final direction of propagation of the CME along the Sun-Earth line at 42 $R_\odot$.
    
\end{enumerate}
Both GCS reconstructions as described by the GCS parameters (Table \ref{tab:indp1} and Table \ref{tab:indp2}), in the Appendix, consistently captured an increasing cross-section of the CME, as represented by the increase in the kappa parameter with time. However, significant degeneracies appeared in the geometric parameters. In fitting 1, the longitude and latitude of the CME flux rope are nearly fixed, showing a radially expanding CME. Fitting 2 shows larger variation, with latitude changing from $–19^\circ$ to $0^\circ$, longitude from $37^\circ$ to $0^\circ$, and tilt from $–24^\circ$ to $–47^\circ$. The differences in the half-angle and tilt for the two fittings can arise due to the tracking of different features by different persons. We can also clearly notice the degeneracy between the heights and the longitudes of two different fittings. Specifically, fitting 2 includes lower heights and higher longitudes of the GCS-fitted CME flux rope up to 20:42 UT, compared to fitting 1, and vice versa beyond this point in time. These differences illustrate the well-known GCS degeneracy, where longitude and height are expected to compensate for each other, and expansion can be represented either as changes in height of the CME flux rope alone or as a combination of changes in height and longitude. This is the manifestation of the projection effects, particularly when the angular separation between two viewpoints is small. Thus, although both fittings agree on the expansion, the geometry remains poorly constrained, underscoring the need for multiple, well-separated viewpoints to reliably determine CME direction and orientation as highlighted by \cite{verbek:2023}.

It is important to note there is an agreement on the initial overall westward propagation of the CME in COR2 FOV (upto $\approx$ 20 $R_\odot$). We found initial longitude of the CME lie between $37^\circ-20^\circ$, from the two fittings at $\approx 5 R_\odot$. This shows an initial westward deflection as compared to the source region location of the CME, i.e., S21W11 as reported by \cite{vemareddy:2024,gopalswamy:2024,pauris:2025}. This can be attributed to the deflection of the CME by a nearby coronal hole in the corona, as suggested by \cite{gopalswamy:2024}.

Based on the above fitting parameters in this work, it is clear that uncertainty arises in the direction of  CME in the HI1 FOV. The range of values of the final longitude obtained from two GCS fittings in HI FOV was 20$^\circ$-0$^\circ$. Therefore, it is unclear if the CME propagated along the Sun-Earth line ($0^\circ$) or $20^\circ$ westward. 

We further explore the implications of different directions of propagation of the April 21, 2023, CME on its kinematics and arrival time estimate. For this purpose, we used the ADBM to estimate the arrival time at different heliocentric distances. We examined the in-situ observations of the CME at different heliocentric distances of different spacecraft and compared them with predictions from the ADBM, corresponding to inputs from different  GCS fittings. 

We would like to note that observation from a third viewpoint (SolO) was available for this CME  \citep{pauris:2025}, who showed that incorporating observations from this additional perspective, in principle, reduced the errors in the GCS fittings and thus provided improved estimates of the CME kinematics. The focus of the present work is to investigate the implications of errors in estimating the ToA of the CME at different spacecraft that are separated by only a small angular distance. We also demonstrate how the two in-situ observations (discussed in the next Section) can be used to better constrain or verify the CME propagation direction in the heliosphere, if the third viewpoint (SolO) had not been available.

\subsection{In-situ observations of the CME}

\begin{figure*}%%[H]
\begin{center}
\includegraphics[width=\linewidth]{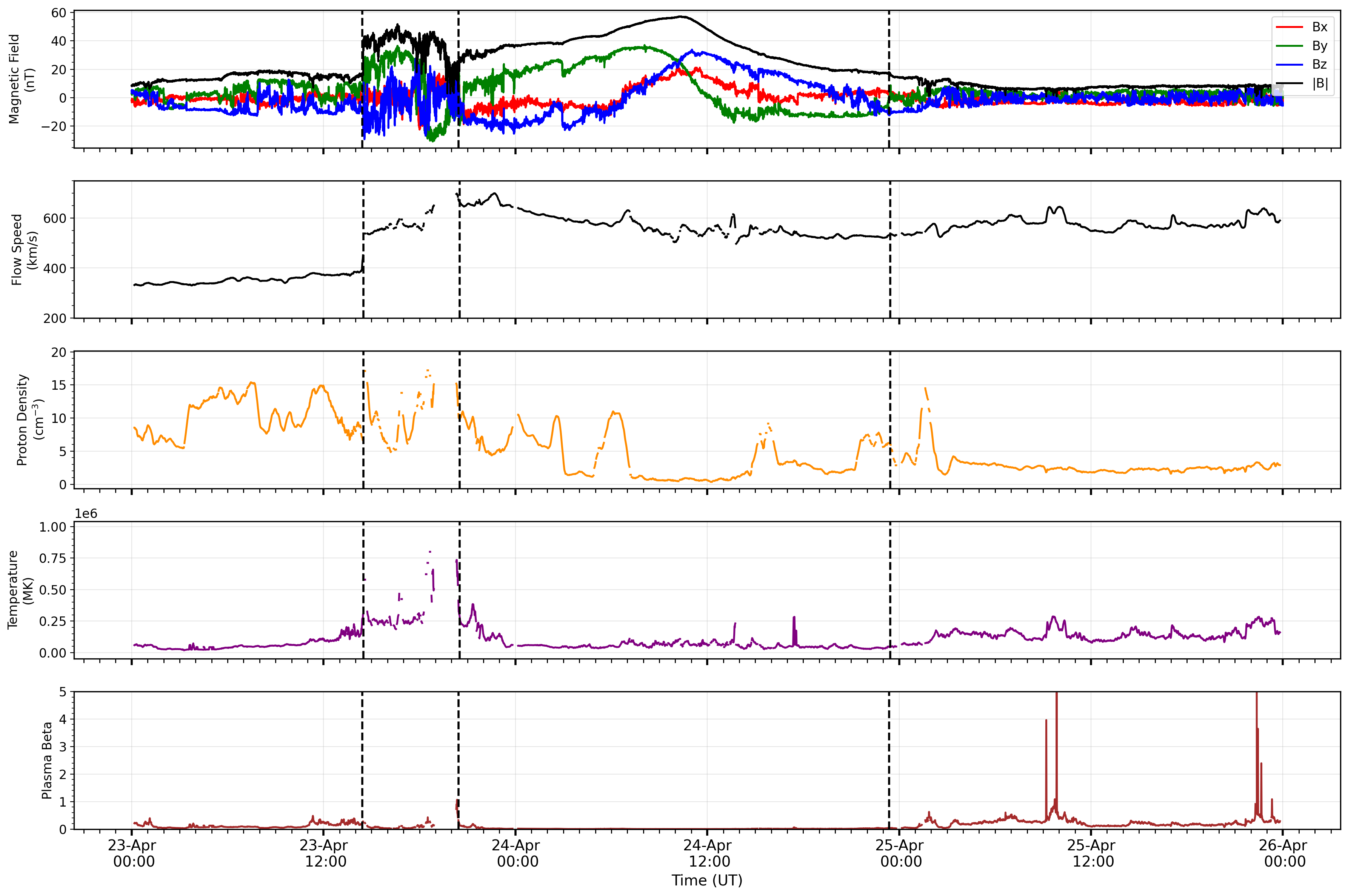}
\caption{In-situ observations by the STEREO-A spacecraft from 23 to 26 April 2023 showing, from top to bottom: the IMF vectors, solar wind flow speed, proton density, and proton temperature in RTN coordinate. The first vertical dashed black line marks the arrival of the CME-driven shock, while the second and third lines indicate the start and end times of the associated magnetic cloud, respectively.
}
\label{fig:insit_STA}
\end{center}
\end{figure*}

\begin{figure*}%%[H]
\begin{center}
\includegraphics[width=\linewidth]{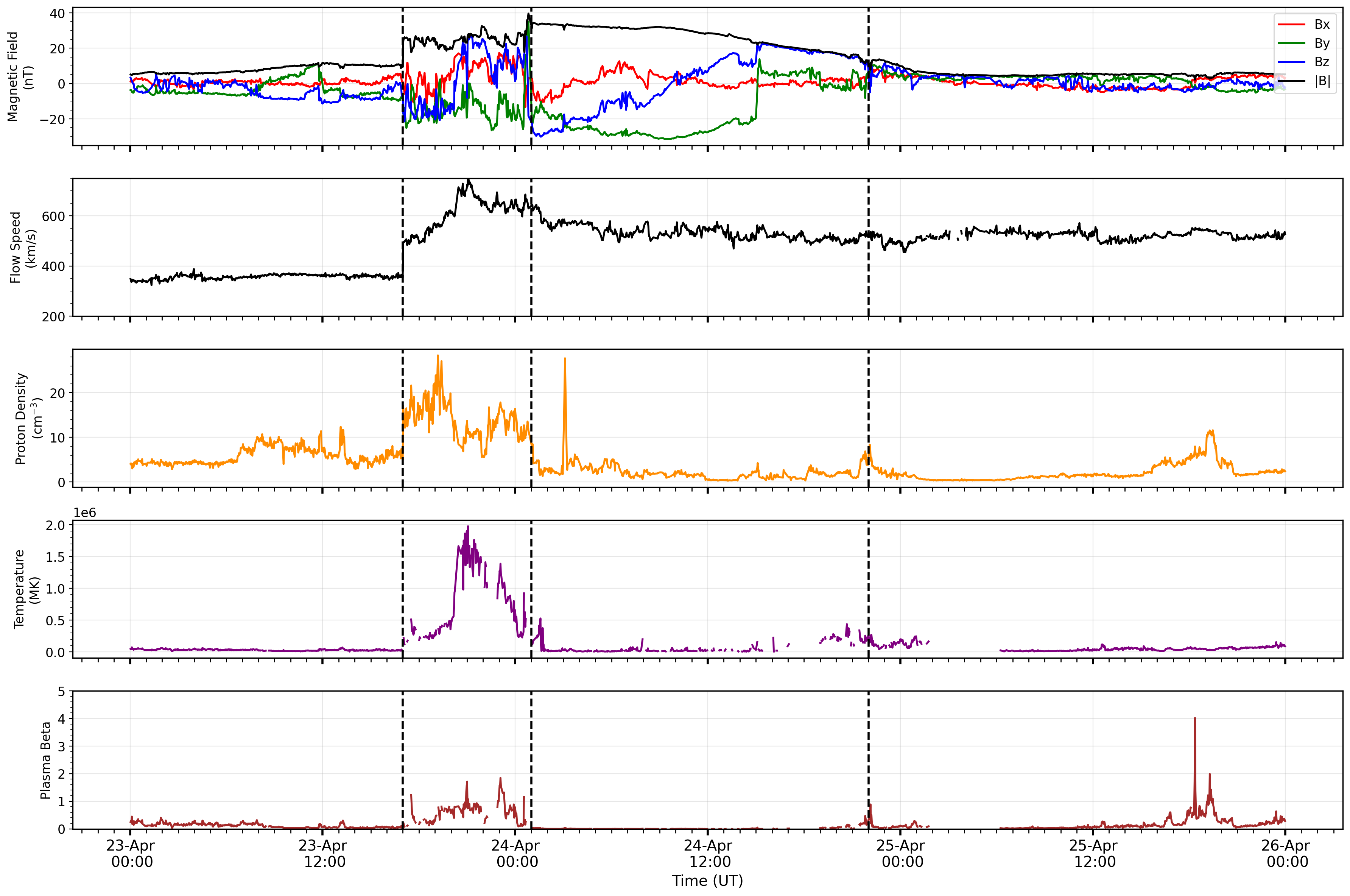}
\caption{ In-situ parameters at L1 observed by the Wind spacecraft from 23 to 26 April 2023. IMF vectors, total magnetic field, plasma flow speed and temperature are plotted with time from top to bottom, respectively in GSE coordinate. The first vertical black line marks the arrival of the shock of the CME, and the second and third dashed lines represent the start and end of the magnetic cloud, respectively.}
\label{fig:insitu_wind}
\end{center}
\end{figure*}

As we have mentioned in Section 2,  the CME passed through three  spacecraft : BepiColombo, STEREO-A, and Wind, respectively. We plotted the in-situ data from STEREO-A and Wind spacecraft in Figure \ref{fig:insit_STA} and \ref{fig:insitu_wind}, respectively. In Figure \ref{fig:insit_STA} and \ref{fig:insitu_wind}, the vertical black lines represent the start time of shock, the start time of the MC, and the end time of the MC, respectively. 
As reported in the HELCATS catalog \footnote{\url{https://helioforecast.space/icmecat/ICME_BEPI_MOESTL_20230422_01}} and Richardson-Cane Catalogue \footnote{(\url{https://izw1.caltech.edu/ACE/ASC/DATA/level3/icmetable2.htm})},  the shock reached BepiColombo on April 22, 2023, at 18:28 UT, which was followed by the arrival of a MC  at the spacecraft on April 23, 2023, at 00:11 UT. The MC ended at 07:32 UT on the same day. Since SolO was located almost opposite to the Sun-Earth line with an angle of $130^\circ$ with Sun-Earth line, the CME did not pass through it.

The CME shock was detected at STEREO-A on April 23, 2023, at 14:29 UT. The MC began at 20:30 UT and ended on April 24, 2023, at 23:25 UT (Figure\ref{fig:insit_STA}).

The Wind spacecraft observed the CME shock on April 23, 2023, 17:02 UT. The MC passage in Wind was from April 24, 2023, at 01:00 UT to 22:02 UT (Figure\ref{fig:insitu_wind}). We have tabulated the shock arrival time, MC start, and end time recorded by different spacecraft in Table \ref{tab:ob_shock_mc}.

These observations reveal that the MC of the CME, after passing through BepiColombo, reached STEREO-A first and arrived at the Wind spacecraft approximately 4 hours and 30 minutes later. We want to emphasize the chronology of the arrival of the CME at STEREO-A and Earth is crucial for comparison with the prediction from ADBM obtained from two GCS fittings.

It is interesting to note here that the speed of the solar wind is fast even after the passage of the CME at L1 as shown in Figure \ref{fig:insitu_wind}. There is a significant increase in the solar wind speed as observed before the shock (17:02 UT, 23 April) and after the CME passage (22:02 UT, 24 April). This suggests a transition from slow ($\approx 350$ km/s) to fast solar wind ($\approx$  650 km/s) often, related to the passage of a  Stream Interacting Region (SIR) around the ICME time \citep{sir:2018}.


\begin{figure*}%%[H]
\begin{center}
\includegraphics[width=\linewidth]{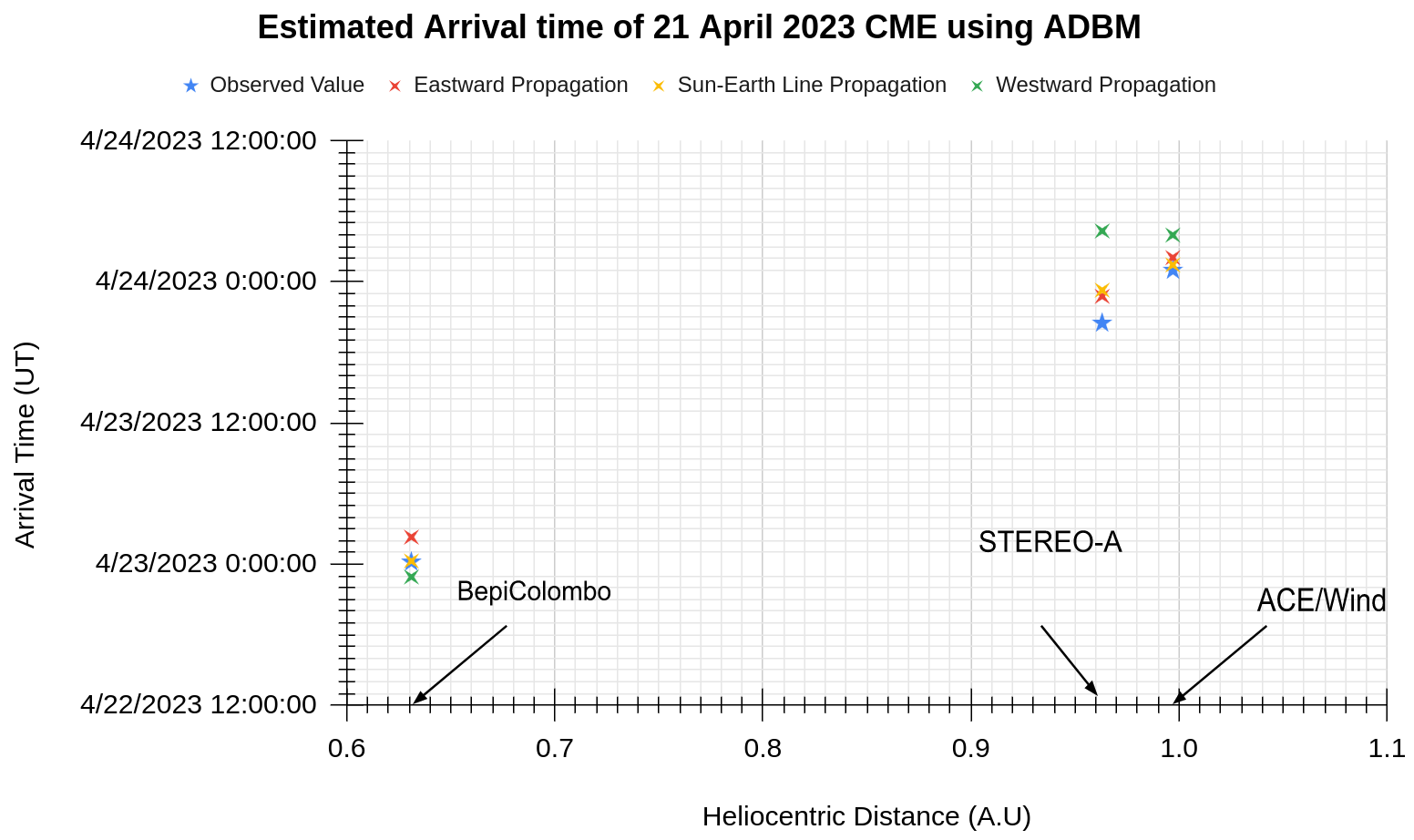}
\caption{ Scatter plot of ToA of MC versus heliocentric distances(AU) of different spacecraft, i.e, BepiColombo, STEREO-A, and Wind, respectively. The blue star represents the observed ToA of the MC at each spacecraft. The red crosses represent the ToA of MC estimated considering the case of eastward propagation where the final longitude reaches -10$^\circ$. The yellow crosses represent the case where the CME is considered to propagate along the Sun-Earth line. The green crosses are the ToA estimated by considering the westward propagation of the CME.} 
\label{fig:ToA}
\end{center}
\end{figure*}
%\newpage

\section{Time of arrival (ToA) estimate based on Advanced Drag-Based Model (ADBM)}
The Advanced Drag-Based Model \footnote{\url{https://oh.geof.unizg.hr/DBM/dbm.php}} (ADBM) was implemented to forecast the transit speed and ToA of the ejecta at each spacecraft based on the results of two  GCS reconstructions. As discussed in the Section 2, two different CME longitude estimates were considered in the ADBM. We then estimated the ToA and impact speed of the CME magnetic cloud (MC) for each scenario and compared these predictions with the actual ToA and impact speeds observed at the respective spacecraft. The input parameters were obtained from the GCS flux rope fitting value of the CME as follows:
\begin{enumerate}
\item CME Take-off Date and time: The date/time of the last image used for the last fitting of the GCS model in HI FOV (April 21, 2023, 23:54 UT) was used as CME take-off time. %Since we have performed GCS reconstruction in HI FOV, this gives an advantage to track the CME  to higher heights as compared to previous studies, which use a typical value of 21 $R_\odot$ \citep{martinic:2023,kumar:2023,kumar:2024}.
\item Starting Radial Distance of the CME: It is the average of the two true heights of the CME obtained from the last GCS fit in HI1 FOV from two  GCS fittings, i.e., approximately 40 $R_\odot$.
\item  Speed of the CME: The true speed of the CME is estimated using the GCS fitting. For this, we considered the average speed from the two GCS fittings performed by the two authors. Its value is 1100 km/s.
\item Drag Parameter: It is a variable that ranges from 0.1 to 2 $\times$ $10^{-7}$. Different values of drag parameters within this range are used to improve the ToA and impact speed accuracy. We used an optimum value of the drag = 2.5$\times10^{-8}$, which provided us with overall better results.
\item Solar Wind Speed: The background solar wind speed is assumed constant and approximately 350 km/s for this CME. This value is taken from the ambient solar wind velocity observed before the CME arrival at L1 (from Wind data).
\item $R_{target}$: Here, the heliospheric distances of the three spacecraft, i.e., BepiColombo, STEREO-A, and Wind, are used (Table \ref{tab:ob_shock_mc}).
\item CME Angular Width: This is the CME half-angular width in the ecliptic plane. We derived this parameter by projecting the GCS CME mesh onto the ecliptic plane.
We used the following formula from spherical trigonometry to calculate the half angular width ($Hf_{ADBM}$) used in the ADBM: $Hf_{\mathrm{ADBM}} = \arctan\!\big( \tan(Hf_{\mathrm{total}})\,\cos(\mathrm{tilt}) \big)$. This is a projection of the CME mesh in the ecliptic plane. Here $Hf_{\mathrm{total}}$= GCS fitted half angle + $\sin^{-1} (kappa)$ and $tilt$ is the GCS fitted tilt angle. %This geometry is shown in Figure \ref{fig:gm}}

\item Earth-target heliocentric angular separation: We have used the angular separation between the Spacecraft and Earth (Table \ref{tab:ob_shock_mc}.).
\item Source region central meridian distance:  This corresponds to the longitude of the CME obtained from GCS reconstruction in the last frame in HI FOV.

\end{enumerate}

%\begin{figure}%%[H]
%\begin{center}
%\includegraphics[width=\linewidth]{Fig_GM.png}
%\includegraphics[width=\linewidth]{Fig_7b.png}
%\caption{Geometry assumed to calculate the half angle ($Hf_{\mathrm{ADBM}}$) for the input of ADBM. Here $Hf_{\mathrm{total}}$= GCS half angle + $\sin^{-1} (kappa)$ and $tilt$ is the GCS fitted tilt angle. }
%\label{fig:gm}
%\end{center}
%\end{figure}

We have listed all the input parameters for ADBM in Table \ref{tab:shock_mc_times} in the Appendix.
As the ADBM estimates the arrival of the MC at different spacecraft, using these parameters mentioned above and two cases of GCS fitting, the ToA and transit speed of the CME at the three spacecraft are determined. These are then compared with the actual ToA of the CME (start time of MC as mentioned in Table \ref{tab:ob_shock_mc}) in the respective spacecraft.
\begin{figure*}%%[H]
\begin{center}
\includegraphics[width=\linewidth]{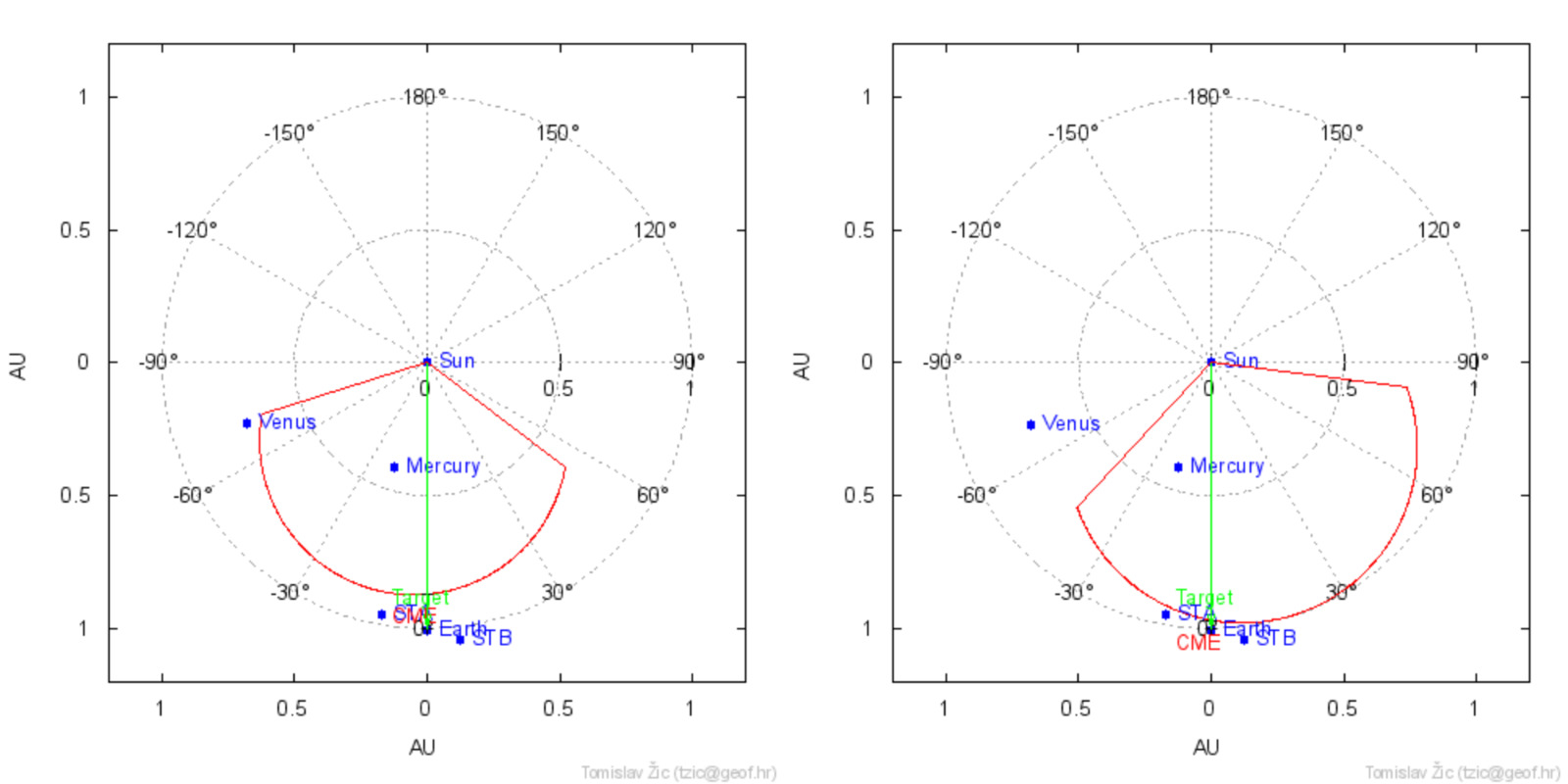}
\caption{CME ejecta arrival at different spacecraft based on the two possible cases of the eastward and westward propagation of the CME. The left panel shows a snapshot of the ADBM output for eastward propagation of CME ejecta, showing the arrival of the CME first at STEREO-A (blue dot on the left), then at Wind spacecraft (green dot). The right panel shows a snapshot of the ADBM output for the case of westward propagation, showing the arrival of the CME ejecta first at Wind and then at the STEREO-A spacecraft. The image/gif was generated by the online tool \protect \footnote{\url{https://oh.geof.unizg.hr/DBM/dbm.php}}.}
\label{fig:adbm_out}
\end{center}
\end{figure*}

In Figure \ref{fig:ToA}, we have plotted the different cases of predicted ToA of CME ejecta  (marked with a cross) and the actual ToA CME ejecta(marked with a star) with the radial position of the spacecraft from the Sun. From Figure\ref{fig:ToA}, Table \ref{tab:adb_shock_mc_times}, and Table \ref{tab:speed_adb}, we can notice that:
\begin{enumerate}
    \item For the case of GCS fitting corresponding to final Earthward propagation (yellow cross) due to eastward deflection from COR2 to HI FOV in GCS fitting, the prediction errors of ToA and transit speed of CME are the least for all three spacecraft. Also, the CME arrives at the STEREO-A spacecraft earlier than the Wind spacecraft.
    \item For the case of GCS fitting, suggesting no deflection in the CME propagation direction, leading to a final $20^\circ$ westward propagation (green cross), the errors in ToA ($\ge 3hr$) and transit speed of CME in STEREO-A is high ($\ge 50$ km/s). Also, the CME  arrives in Wind spacecraft and then at STEREO-A.   
\end{enumerate}
In the case of Earthward propagation, the nose of the CME is heading toward the STEREO-A spacecraft, and thus, the CME arrives at STEREO-A first, then at the Wind spacecraft. In the case of $20^\circ$ westward propagation, the flank part of the CME  arrives at the Wind and then at STEREO-A spacecraft. 

It is important to note that in the case of $20^\circ$ westward propagation of the CME, the error (+7:50 hr) in ToA are larger in the STEREO-A spacecraft as compared to Earthward propagation (+2:48 hr). Moreover,  ADBM output predicts that CME first arrived at Wind at 3:59 UT on 24 April and then at STEREO-A at 4:20 UT on 24 April, contrary to the actual observations. 
The chronology of the arrival of the CME at STEREO-A and Wind matches only in the case of Earthward propagation of the CME. In this case, CME first arrives at STEREO-A at 23:18 UT (+2:48 hrs of actual ToA) on 23 April and then at the Earth at 01:27 UT on 24 April (00:27 hr of actual ToA). Moreover, the difference between the actual and predicted ToA for BepiColombo is only 5 minutes.
The GCS fitting corresponding to this case suggested the eastward deflection from its initial/first direction of propagation. i.e., the direction of propagation at the height above approx 40 $R_\odot$ (insertion time of the input to ADBM) is Earth-directed (stonyhurst longitude 0$^\circ$). This suggests a net eastward deflection of the CME from its initial direction of propagation, i.e., in COR2 FOV.
Motivated by the above finding, we increased the input longitude of the CME in the eastward direction in the ADBM to improve the ToA prediction of the CME at STEREO-A and Wind. We used longitude -10$^\circ$ (same as the longitude of the STEREO-A), with speed = 1100 km/s, the starting distance as 40 $R_\odot$, and the starting time as 23:54 UT, and the rest of the parameters are the same as the previous two cases. The corresponding values of ToA are plotted in Figure\ref{fig:ToA} with red crosses and tabulated in Table \ref{tab:adb_shock_mc_times}.
For this case, the arrival of CME at STEREO-A was on 23 April at 22:47 UT with an error of +02:17 hr with actual ToA. Whereas predicted arrival of the CME at Wind for this case was on  24 April at 02:04 UT with an error of +01:04 hr with the actual ToA at Wind. Moreover, this resulted in a difference of ToA for STEREO-A and Wind, approx 03:30 hr, which is close to the actual difference between ToA, i.e., 04:30 hr, at these two spacecraft.
Therefore, using input longitude -10$^\circ$ in ADBM (nearly the exact longitude as STEREO-A) gave us the best results with respect to the chronology of CME arrival at STEREO-A and Earth. 
Thus, using the ADBM, we inferred that the CME underwent an eastward deflection during its propagation from near the Sun  (COR2 FOV) to the heliosphere (HI FOV). In the heliosphere final direction propagation is along the STEREO-A-Sun line.
We want to mention that due to the assumed shape of the CME in ADBM and the relative location of STEREO-A and Wind, the chronology of the arrival of the CME at STEREO-A and Wind depends only upon the longitude of the CME in the ADBM. It is also possible that by using different inputs for speed and height estimated from GCS for each GCS fitting to ADBM, slightly different results can be obtained for individual spacecraft; however, in this work, we only focused on the ADBM parameter that differentiates the chronology of the arrival of the CME at STEREO-A and Wind, i.e., the longitude of the CME. Therefore, in the above analysis, we only discussed the results corresponding to different longitudes of the two GCS fittings and used average values for other ADBM parameters.
\section{Discussion}
This study based on the 3D reconstruction of the CME on 21 April 2023 close to the Sun and estimation of the ToA using the ADBM, suggests that the CME, which originated at S21W11, underwent an overall eastward deflection while propagating in the heliosphere till 1 AU.
We arrived at this conclusion by comparing the ADBM predictions of ToA with the actual ToA of CME at different spacecraft located at different heliocentric distances, i.e., BepiColombo (0.631 AU), STEREO-A (0.963 AU) and Wind (0.997 AU).

Our analysis highlights the broader importance of using two-point in-situ observations to reduce the directional ambiguity. When only limited viewpoint observations with small separation angles are available, combining ADBM analysis with two-point in-situ observations can substantially mitigate these errors. Thus, our approach provides a valuable methodology for similar cases to reduce the errors in the estimation of the direction of propagation of the CME.
Recent analysis of this CME by \cite{pauris:2025} demonstrated that incorporating three-viewpoint observations (SOHO, STEREO-A, and Solar Orbiter) into the 3D reconstruction significantly improved the accuracy of CME kinematic and geometric parameters, as well as the ToA at Earth. In contrast, in the present study we consider the chronology of its arrival at the different spacecraft located at different heliocentric distances, i.e., STEREO-A (0.963 AU) and Wind (0.997 AU). This enabled us to reach a robust conclusion regarding its propagation direction without even incorporating a third view-point. Our conclusion about the Earth-directed propagation of the CME agrees with the analysis of \cite{pauris:2025}, which reported only a $5^\circ$ separation between the CME propagation and Sun-Earth line in the heliosphere. In fact, \cite{pauris:2025} also showed that GCS fitting, using only SOHO and STEREO viewpoints does not differ significantly from the case when Solo observations were also incorporated, up to 22 April 00:00 UT. This time corresponds to the last fitting in our case, i.e., at 21 April, 23:54. Moreover for their ToA estimates \cite{pauris:2025} assumed a constant speed of the CME after the CME tracking in the heliosphere. However, ADBM  employed in our analysis consider the CME acceleration/deceleration throughout the heliosphere due to CME-solar wind interaction.
The possible cause of the CME deflection in the heliosphere is also a subject of research that influences the propagation. It should also be noted that the drag-based model employed in our analysis assumes a semi-circular CME front. However, this assumption may not hold as reported in earlier studies, where CMEs are observed to undergo front flattening, distortions, and other large-scale deformations \cite{Savani:2010,Owens:2017,Chi:2021,yang:2023}.
 Therefore, this needs to be explored using full 3D MHD modelling in the heliosphere. 
This eastward deflection can range from a maximum of E10$^\circ$ (20$^\circ$ East, from the source region) to the Sun-Earth line (11$^\circ$ West, from the source region). This happens to be the most favorable scenario for explaining the arrival of the CME in STEREO-A  first, then at Wind spacecraft.

We note that \cite{gopalswamy:2024} suggested that assuming a more westward propagation direction ($\approx 40^\circ$) decreases the estimated CME speed in the empirical relation, leading to a longer transit time and reduced errors ($\approx 2 hr$). They suggested the reason for this westward deflection could be the presence of a nearby coronal hole. We extend the analysis to investigate the sequence of the CME arrival at STEREO-A and Wind, which strongly hints towards an eastward deflection of the 21 April 2023 CME in the heliosphere.

It is important to mention here that due to uncertainty in the GCS parameters and different inputs of the ADBM, one could easily tweak the  ADBM parameters (speed and direction) to find the best arrival estimate closely matching with the observations, individually for Wind or at STEREO, even in the case of the westward direction of propagation of the CME. However, due to the assumed shape of the CME in the ADBM (circular front), the relative position of the spacecraft, and relative heliocentric distances, it is impossible to obtain the chronology correct for the spacecraft simultaneously for the same input parameter for westward (longitude $\ge 10 ^\circ$ in West of Sun-Earth line) propagating CME.\\

Our final conclusion about the eastward deflection of this fast CME in the heliosphere agrees with previous studies by \cite{Wang:2004,Wang:2014,kumar:2024}. \cite{Wang:2004} reported that a fast CME of speed $\approx$ 1100 km/s, while propagating in a slow solar wind of speed $\approx$ 450 km/s (Figure: 7 of \cite{Wang:2004}), tends to deflect in an eastward direction by $\approx25^\circ$. This value is comparable with the range of values we observed, i.e., $\approx 20^\circ-30^\circ$.

This study suggests that using individual simplistic models for CME propagation and tracking might not be sufficient to conclude about the CME propagation direction, mainly when the separation angle between two observing spacecraft is small. It has further implications when used as an input in another model, i.e., ADBM, leading to errors in ToA. Therefore, a more conclusive finding can be made by reconciling the prediction from different techniques (GCS and ADBM) along with two point situ observations (STEREO-A, Wind, and BepiColombo) to constrain the CME propagation direction.

%The in-situ observations of Wind and STEREO-A suggest the passage of a sector boundary with a substantial solar wind speed, which provides us with a possibility of the CME shock interacting with the SIR. Further investigation and analysis are required to establish the CME-SIR interactions. 

\section*{Acknowledgements}
We acknowledge the use of Wind and STEREO in-situ data from the OMNI database using pySPEDAS \citep{pyspd:2024}. We acknowledge the use of the GCS code \citep{gcs:2024} and ADBM\footnote{\url{https://oh.geof.unizg.hr/DBM/dbm.php}}. We acknowledge the Sunpy community\citep{Sunpy_community2020}. We acknowledge the HELCATS catalogue \citep{hel:2017} and Richardsen Cane Catalogue \citep{rich:2010}.
Coronagraphic images are used from Helioviewer and HI1 images from the site (\url{https://stereo-ssc.nascom.nasa.gov/data/ins_data/secchi/secchi_hi/L2_11_25/}). This work was carried out under the Indo-U.S. Science and Technology Forum (IUSSTF) Virtual Network Center project (Ref. no. IUSSTF/JC-113/2019). Parthib Banerjee acknowledges the opportunity provided by PRL to carry out his master's project at USO/PRL. NG is supported by NASA's STEREO project and the Living With a Star program. 

\newpage
\begin{table*}[ht]
\caption{Shock and MC arrival times at different spacecraft (Spc) along with their heliocentric locations in HEE coordinates.\label{tab:ob_shock_mc}. Here, BpC and STA stand for BepiColombo and STEREO-A, respectively. Shock arrival time and MC interval are provided in UT.} 
\renewcommand{\arraystretch}{1.1}
\begin{tabular}{|c|c|c|c|}
\hline
\textbf{Spc} & \textbf{Location (R, $\phi$, $\theta$)} & \textbf{Shock Time} & \textbf{MC Interval} \\
\hline
BpC & 0.631 AU, 21.9$^\circ$, -3.4$^\circ$ & 22 Apr 18:28 & 23 Apr 00:11 -- 07:32 \\
\hline
STA & 0.963 AU, -10.2$^\circ$, -5.8$^\circ$ & 23 Apr 14:29 & 23 Apr 20:30 -- 24 Apr 23:25 \\
\hline
Wind & 0.997 AU, -0.1$^\circ$, -4.9$^\circ$ & 23 Apr 17:02 & 24 Apr 01:00 -- 22:02 \\
\hline
\end{tabular}
\end{table*}

\begin{table*}

\caption{ADBM CME parameters used in the analysis of the 21 April 2023 event.}
\label{tab:shock_mc_times}
\begin{tabular}{|l|c|c|c|}
        \hline
        \textbf{ADBM parameters} & \textbf{EW} & \textbf{ERW} & \textbf{WW} \\
        \hline
        Takeoff Time (April 21, 2023) & 23:54 UT & 23:54 UT & 23:54 UT\\
        \hline
        Starting Radial Distance ($R_\odot$) & 40 & 40 & 40 \\
        \hline
        Speed of the CME (km/s) & 1100  & 1100 & 1100  \\
        \hline
        Drag Parameter & 0.25 & 0.25 & 0.25 \\
        \hline
        Solar Wind Speed (km/s)& 350 & 350 & 350 \\
        \hline
        $R_{target}$ & SC distance & SC distance & SC distance  \\
        \hline
        CME Angular Width ($^\circ$) & 63 & 66 & 63 \\
        \hline
        Earth Directed heliocentric angular separation ($^\circ$) & SC $\phi$  & SC $\phi$ &  SC $\phi$ \\
        \hline
        Source Region Central Meridian Distance ($^\circ$) & -10 & 0 & 22 \\
        \hline
\end{tabular}
\end{table*}

\begin{table*}
\caption{ Comparison of arrival time estimate of different cases of the direction of propagation of predicted ToA with the observed ToA of MC at respective spacecraft. The values in brackets represent the errors of the different cases (hh:mm). Here, EW, ERW, and WW show the cases of eastward propagation, Earthward (Sun-Earth line) propagation, and westward propagation of CME, respectively.}
\label{tab:adb_shock_mc_times}
\begin{tabular}{|l|c|c|c|c|}
         \hline
        \textbf{Spacecraft} & \textbf{Obs. ToA } & \textbf{EW ($\phi=$-10$^
        \circ$)} & \textbf{ERW($\phi$=0$^
        \circ$)} & \textbf{WW ($\phi=$ 20$^
        \circ$ )}  \\
        \hline
        BepC  & 23 April 0:11 & 23 April 2:18(+(2:07)) & 23 April 00:16 (+00:05) & 23 April 22:55(-01:16) \\
        \hline
        STEREO-A  & 23 April 20:30 & 23 April 22:47 (+2:17) & 24 April 23:18 (+2:48) & 24 April 4:20(+7:50) \\
        \hline
    Wind & 24 April 1:00 & 24 April 2:04 (+01:04) & 24 April 1:27 (00:27) & 24 April 3:59(+2:59)\\
        \hline

\end{tabular}
\end{table*}

\begin{table*}
\caption{Comparison of predicted and observed CME transit speeds under different propagation scenarios. Values in brackets indicate errors. No transit speed is listed for BepiColombo due to missing solar wind data.}
\label{tab:speed_adb}

\begin{tabular}{|l|c|c|c|c|c|c|}
          \hline
        \textbf{SC} & \textbf{Actual Transit Speed (km/s)} & \textbf{EW($\phi$=-10) (km/s)} & \textbf{ERW($\phi$=0) (km/s)} & \textbf{WW (km/s)}  \\
        \hline
        BepiColombo & No Data & 575 & 610 & 640\\
        \hline
        STEREO-A& 550 & 530(-20) & 525(-25) & 480(-70)  \\
        \hline
        Wind  & 550 & 520(-30) & 525(-25) & 500(-50) \\
        \hline

\end{tabular}
\end{table*}

\bibliography{manu.bib}
\appendix

\vspace{2cm}
\begin{table*}
\caption{GCS parameters estimated from first fitting. }
\label{tab:indp1}
\begin{tabular}{|c|c|c|c|c|c|c|c|c|}
\hline
\textbf{Date} & \textbf{Time (UT)} & \textbf{Half Angle} & \textbf{Kappa} & \textbf{Latitude} & \textbf{Longitude} & \textbf{Tilt} & \textbf{Height}  \\ \hline
2023-04-21    & 18:24  & 28  & 0.61 & -15 & 20 &-26 & 5.7     \\ \hline
2023-04-21    & 18:54  & 28  & 0.61 & -16 & 20 &-26 & 9.8   \\ \hline
2023-04-21    & 19:18  & 28  & 0.61 & -16 & 20 &-28 & 12.2  \\ \hline
2023-04-21    & 19:54  & 28  & 0.61 & -19 & 20 &-28 & 16.8  \\ \hline
2023-04-21    & 20:18 & 28 & 0.61 & -16 & 20 &-28 & 18.4   \\ \hline
2023-04-21    & 20:42 & 28  & 0.62 & -17 & 20 &-29 & 20.4 \\ \hline
2023-04-21    & 21:18 & 28  & 0.62 & -14 & 20 &-29 & 23.9  \\ \hline
2023-04-21    & 21:54 & 28  & 0.62 & -14 & 20 &-29 & 27.0   \\ \hline
2023-04-21    & 22:18 & 28  & 0.62 & -15 & 20 &-29 & 30.0  \\ \hline
2023-04-21    & 23:06 & 28  & 0.62 & -15 & 22 &-30 & 33.0 \\ \hline
2023-04-21    & 23:18 & 28  & 0.62 & -15 & 22 &-30 & 34.5 \\ \hline
2023-04-21    & 23:54 & 28  & 0.62 & -15 & 22 &-30 & 38.0 \\ \hline
\end{tabular}
\end{table*}
\begin{table*}
\caption{GCS parameters estimated second fitting.}
\label{tab:indp2}
\begin{tabular}{|c|c|c|c|c|c|c|c|c|}
\hline
\textbf{Date} & \textbf{Time (UT)} & \textbf{Half Angle} & \textbf{Kappa} & \textbf{Latitude} & \textbf{Longitude} & \textbf{Tilt} & \textbf{Height}  \\ \hline
2023-04-21    & 18:24  & 31  & 0.61 & -19 & 37 &-24 & 4.7     \\ \hline
2023-04-21    & 18:54  & 34  & 0.61 & -18 & 37 &-34 & 8.0     \\ \hline
2023-04-21    & 19:18  & 34  & 0.62 & -21 & 34 &-47 & 11.4   \\ \hline
2023-04-21    & 19:54  & 34  & 0.62 & -18 & 30 &-47 & 14.9\\ \hline
2023-04-21    & 20:18 & 34  & 0.62 & -16 & 25 &-47 & 16.8 \\ \hline
2023-04-21    & 20:42 & 34  & 0.62 & -17 & 24 &-47 & 20.0 \\ \hline
%2023-04-21    & 20:53  & 38  & 0.65 & -13 & 26 &-49 & 20  \\ \hline
2023-04-21    & 21:18  & 34  & 0.62 & -11 & 15 &-47 & 26  \\ \hline
2023-04-21    & 21:54 & 34  & 0.63 & -13 & 11 &-47 & 28   \\ \hline
2023-04-21    & 22:18 & 34  & 0.64 & -11 & 10 &-47 & 31.0  \\ \hline
2023-04-21    & 23:06 & 34  & 0.64 & -7 & 7 &-47& 36 \\ \hline
2023-04-21    & 23:18 & 34  & 0.64 & -2 & 6 &-47 & 37 \\ \hline
2023-04-21    & 23:54 & 34  & 0.64 & 0 & 0 &-47 & 42.0\\ \hline
\end{tabular}
\end{table*}

\end{document}